\documentclass[epj]{svjour}
\usepackage{epsfig} 
\usepackage{times}
\begin{document}
%
%
\title{{\boldmath $\phi$}-meson photoproduction from nuclei}
\author{A.~Sibirtsev\inst{1}, H.-W.~Hammer\inst{1},
U.-G.~Mei{\ss}ner\inst{1,2} and A.W. Thomas\inst{3}} 
\institute{Helmholtz-Institut f\"ur Strahlen- und Kernphysik (Theorie), 
Universit\"at Bonn, Nu\ss allee 14-16, D-53115 Bonn, Germany \and
Institut f\"ur Kernphysik (Theorie), Forschungszentrum J\"ulich,
D-52425 J\"ulich, Germany \and Jefferson Lab, 12000 Jefferson Ave., 
Newport News, VA 23606, USA}
\date{Received: date / Revised version: date}

\abstract{We study coherent and incoherent $\phi$-meson
photoproduction from nuclei. The available data are analyzed in terms
of single and coupled channel photoproduction. It is found that the data on
coherent photoproduction can be well reproduced within a single
channel optical model and show only little room for 
$\omega-\phi$ mixing. These data indicate a normal distortion of the
$\phi$-meson in nuclei, which is compatible with the results obtained
through the vector meson dominance model. The data on incoherent
$\phi$-meson photoproduction show an anomalous $A$-dependence resulting
in a very strong $\phi$-meson distortion. These data can be explained by
a coupled channel effect through the dominant contribution from the
$\omega\to\phi$ or $\pi{\to}\phi$ transition or, more speculative,
through the excitation of a cryptoexotic $B_\phi$-baryon.}  


\authorrunning{A.~Sibirtsev, H.-W.~Hammer, U.-G.~Mei{\ss}ner, A.W.~Thomas}
\titlerunning{${\rm\phi}$-meson photoproduction off nuclei.}

\maketitle
\section{Introduction}
The renormalization of the  meson spectral function in nuclear
matter (for some early references, see
\cite{Hatsuda1,Bernard1,Bernard2,Brown,Hatsuda2,Ko} and a recent 
review is~\cite{Saito})
attracted substantial interest in connection with  the 
measurements of the di-lepton invariant mass spectra from heavy ion
collisions~\cite{Agakishev1,Masera,Agakishev2}. 
Recently  experimental results on
$\omega$ and $\phi$ meson modification at normal nuclear densities 
have been reported in experiments involving photon and proton 
beams \cite{Ozawa,Trnka,Naruki,Muto,Tabaru,Lolos}. The most remarkable
result obtained in all these experiments is the anomalous $A$-dependence
of the $\phi$-meson production from nuclear targets. At the same time
the $A$-dependence of the $\omega$-meson production both in $\gamma{A}$ and
$p{A}$ interactions can be well understood. 

At the threshold of elementary $\phi$-meson production, its
momentum in the laboratory system, {\it i.e.} with respect to nuclear
matter, is quite high and a substantial fraction of the $\phi$-mesons decay
outside the nucleus. Only that small fraction which decays inside the
nucleus would indicate a probable pole shift of the spectral function. 
Therefore one could not
expect to  observe  a significant in-medium modification of the
$\phi$-meson  mass by measuring the di-leptonic or $K^+K^-$ invariant mass
spectra. However, it is very plausible to study the modification of the
$\phi$-meson width, since at low densities it is related to the imaginary
part of the forward $\phi{N}$ scattering 
amplitude~\cite{Lenz,Dover,Friman1}. The latter determines the 
$\phi$-meson distortion in the nucleus, which can be studied by
measuring the $A$-dependence of the $\phi$-meson production. 

Such ideas motivated experiments on $\phi$-meson production from nuclei
at the KEK-PS~\cite{Tabaru}, SPRING-8 \cite{Ishikawa} and at
COSY \cite{Hartmann1}. Here we analyze recent results on incoherent
$\phi$-meson\,  photoproduction from \, nuclei \, collected at
SPRING-8~\cite{Ishikawa}, which indicate a substantial distortion of
the $\phi$-meson in finite nuclei. For consistency, we also
analyze data on coherent $\phi$-meson photoproduction collected at
Cornell some time ago \cite{McClellan1}. We investigate the role of single and
coupled  channel effects in $\phi$-meson photoproduction and provide
a possible explanation  of the observed anomaly.

The manuscript is organized as follows. In Sec.~2, we analyze the elementary
amplitudes $\gamma p \to \phi p$ and $\gamma p \to \omega p$ in terms of the
vector meson dominance model and compare to data taken at Cornell, ELSA and
SPRING-8. In Sec.~3, we study coherent and incoherent $\phi$-meson
photoproduction off nuclear targets in a single channel optical 
model and show that such an
approach is not capable of describing the new data from SPRING-8. Coupled
channel scattering is considered in Sec.~4 and it is in particular shown that
the $A$-dependence of the SPRING-8 data can be understood in a two-step model,
including $\omega-\phi$ mixing and coupling to an intermediate pion. We also
add some speculations about the excitation of crypto-exotuic baryons with
hidden strangeness. Sec.~5 contains our conclusions.

\section{Vector Dominance Model}
\subsection{Data evaluation}
Considering quark-anti-quark fluctuations of the photon, the Vector
Dominance Model (VDM) assumes that intermediate hadronic $q{\bar q}$
states are entirely dominated by the neutral vector mesons. In that
sense the hadron-like photon~\cite{Stodolsky}
is a  superposition of all possible vector meson
states. The $\gamma{N}{\to}\phi{N}$ reaction can be decomposed into the
transition of the photon to a virtual vector meson $V$ followed by the
elastic or inelastic vector meson scattering on the target
nucleon. The invariant reaction amplitude follows as~\cite{Bauer,Paul}
\begin{eqnarray}
{\cal M}_{\gamma{N}{\to}\phi{N}} = \sum_{V}\frac{\sqrt{\pi\alpha}}{\gamma_V}
{\cal M}_{VN{\to}\phi N},
\label{ampli1}
\end{eqnarray}
where the summation is performed over vector meson states, $\alpha$ is
the fine structure constant, $\gamma_V$ is the photon coupling to
the vector meson $V$ and ${\cal M}_{VN{\to}\phi N}$ is the amplitude for the
$VN{\to}\phi{N}$ transition.

\begin{figure}[t]
\vspace*{-6mm}
\centerline{\hspace*{5mm}\psfig{file=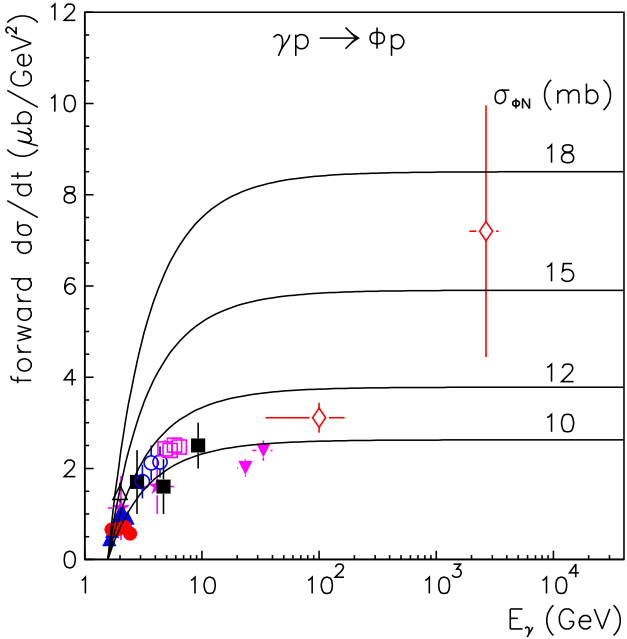,width=9.5cm,height=9.4cm}}
\vspace*{-2mm}
\caption{The forward $\gamma{p}{\to}\phi{p}$ differential cross
section as a function of photon energy. The data are taken from 
Refs.\cite{Erbe,Ballam,Besch,Behrend,Barber,Atkinson,Busenitz,Derrick,Barth,Mibe}.
The lines show the calculations using Eq.(\ref{vdm1}) with $\alpha_\phi$=0
and for different values of $\sigma_{\phi{N}}$.}
\label{spring2}
\end{figure}

A direct determination of  $\gamma_V$ is possible through 
vector meson decay into a lepton pair~\cite{Nambu}
\begin{eqnarray}
\Gamma(V{\to}l^+l^-)=\frac{\pi\alpha^2}{3\gamma_V^2}\sqrt{m_V^2-4m_l^2}
\left[1+\frac{2m_l^2}{m_V^2}\right],
\label{coupl1}
\end{eqnarray}
where $m_V$ and $m_l$ are the masses of vector meson and lepton, respectively.
Taking the di-electron decay width from Ref.~\cite{PDG}, the photon coupling to
the lightest vector mesons is given as
\begin{eqnarray}
\gamma_\rho{\div}\gamma_\omega{\div}\gamma_\phi =2.48{\div}8.53{\div}6.69.
\label{coupl2}
\end{eqnarray}
Although the couplings $\gamma_V$ can be determined
experimentally,  Eq.({\ref{ampli1}) does not indicate whether
non-diagonal, {\it i.e.} $\rho{N}{\to}\phi{N}$ and
$\omega{N}{\to}\phi{N}$, or diagonal $\phi{N}{\to}\phi{N}$ processes
dominate the $\phi$-meson photoproduction on the nucleon, since the
${\cal M}_{V\phi}$ amplitudes can not be measured. Furthermore, VDM
suggests that the virtual vector meson stemming from the photon becomes
real through the four-momentum $t$ transferred to the nucleon, which in
general requires the introduction of a form-factor in the interaction
vertices~\cite{Hufner,Sibirtsev1,Sibirtsev2}. 

\begin{figure}[t]
\vspace*{-6mm}
\centerline{\hspace*{5mm}\psfig{file=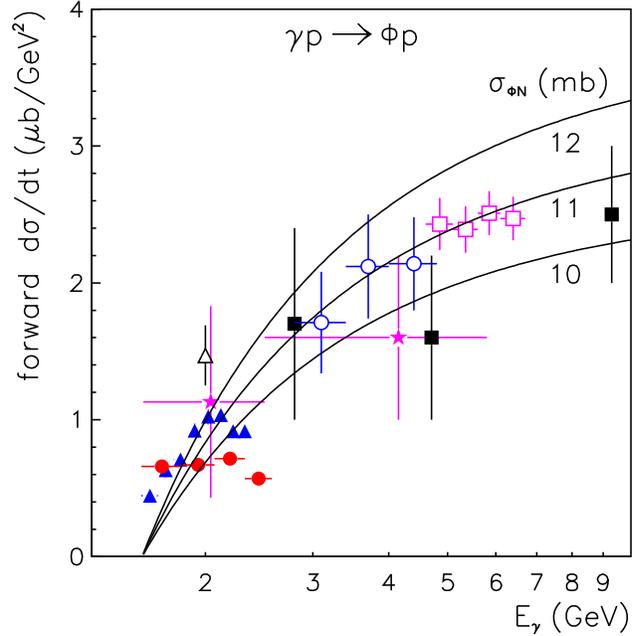,width=9.5cm,height=9.4cm}}
\vspace*{-2mm}
\caption{Same as in Fig.\ref{spring2} for a low photon energy scale. The
solid circles show the results collected by SAPHIR~\cite{Barth}, while
the solid triangles are the measurements from SPRING-8~\cite{Mibe}.}
\label{spring2a}
\end{figure}

The next step in the VDM analysis is to consider only the diagonal
transition or elastic $V{N}{\to}V{N}$ scattering. The imaginary
part of the amplitude $f^\ast_{\phi{N}}(0)$ in the center of mass for 
forward elastic  $\phi{N}{\to}\phi{N}$ scattering
is related to the $\phi{N}$ total cross section, $\sigma_{\phi{N}}$, by an
optical theorem as\footnote{It is clear that this formalism  holds for the
photoproduction of any vector meson $V{=}\rho$, $\omega$, $\phi$,
$J/\Psi,\ldots\,$.}
\begin{eqnarray} 
\Im f^\ast_{\phi{N}{\to}\phi N}(q_\phi, \theta{=}0)=
\frac{q_\phi}{4\pi}\, \sigma_{\phi{N}},
\label{optic}
\end{eqnarray}
where $q_\phi$ and $\theta$ are the $\phi$-meson momentum 
and scattering angle in the $\phi{N}$ center of mass system, 
respectively. The amplitude $f^\ast_{\phi{N}}$ is related to the Lorentz 
invariant scattering amplitude as
\begin{eqnarray} 
{\cal M}_{\phi N{\to}\phi N}=-8\pi\,\, \sqrt{s}\,\, 
f^\ast_{\phi{N}{\to}\phi N}(q_\phi, \theta),
\end{eqnarray}
with $s$ the invariant collision energy squared.
The $\gamma{N}{\to}\phi{N}$ differential cross section is
given in terms of the Lorentz invariant amplitude as
\begin{eqnarray}  
\frac{d\sigma_{\gamma N{\to}\phi N}}{dt}=
\frac{|{\cal M}_{\gamma N{\to}\phi N}|^2}
{64\pi s q_\gamma^2}, 
\end{eqnarray}
where $q_\gamma$ is the photon momentum in the center of mass system. By
introducing the ratio of the real to the imaginary part of the forward
$\phi{N}$ scattering amplitude as $\alpha_\phi$, the
$\gamma{N}{\to}\phi{N}$ differential cross section at $t{=}0$ can be
written as\footnote{Note that the ${q^2_V}/{q^2_\gamma}$ ratio
was not included in most of the VDM analyses reviewed in
Ref.\cite{Bauer}. While this simplification might be applicable at high
energies, that is not the case for most of the data available 
at $E_\gamma{<}10$~GeV, as is illustrated by Fig.\ref{spring2a}.} 
\begin{eqnarray}  
\left.{\frac{d\sigma_{\gamma N{\to}\phi N}}{dt}}\right|_{t{=}0}\!\!\!\!\!
=\frac{\alpha}{16\gamma_\phi^2} \, \frac{q^2_\phi}{q^2_\gamma}\,
(1+\alpha_\phi^2)\, \sigma_{\phi{N}}^2.
\label{vdm1}
\end{eqnarray}
Again, both the ratio $\alpha_\phi$ and the total $\phi{N}$ cross 
section are unknown.
Thus VDM analysis of photoproduction data requires additional
assumptions. 

It is believed that at high energies the hadronic forward scattering amplitudes
are purely imaginary. The data available \cite{PDG} for the $pp$, ${\bar p}p$,
$\pi^-p$, $\pi^+p$, $K^+p$ and $K^-p$ reactions indicate that the
ratios of the real
to imaginary parts of the forward scattering amplitudes are $\simeq$0.1
at $\sqrt{s}{>}$20~GeV. The $\alpha_\phi$ ratio was
measured~\cite{Alvensleben} through the interference between the
$\phi{\to}e^+e^-$ decay and the Bethe-Heitler production of
electron-positron pairs and it was found that
$\alpha_\phi{=}{-}0.48^{{+}0.33}_{{-}0.45}$ at photon energies
6${<}E_\gamma{<}$7.4~GeV. This result is too uncertain to be used
in the further analysis. To estimate the maximum value of
$\sigma_{\phi{N}}$, we apply $\alpha_\phi{=}0$. It is clear that any
non--vanishing ratio $\alpha_\phi$ would result in a reduction of the
$\phi{N}$ cross section that will be evaluated from the data in what follows.

Now Fig.~\ref{spring2} shows the available 
data~\cite{Erbe,Ballam,Besch,Behrend,Barber,Atkinson,Busenitz,Derrick,Barth,Mibe} 
on the forward $\gamma{p}{\to}\phi{p}$ differential cross section as a
function of the photon energy. The lines are the results using
Eq.~(\ref{vdm1}) obtained with $\alpha_\phi$=0 and for different values
of the total $\phi{N}$ cross section. Only one experimental point at
high energy~\cite{Derrick} needs a large  $\sigma_{\phi{N}}$, although
within experimental uncertainty this measurement is in agreement with
the data at photon energies below 10~GeV, which are also shown 
in Fig.~\ref{spring2a}. Note that a large part of the data shown in
Fig.~\ref{spring2a} is in reasonable agreement with calculations done with
$\sigma_{\phi{N}}{\simeq}$11~mb. 

\begin{figure}[t]
\vspace*{-7mm}
\centerline{\hspace*{5mm}\psfig{file=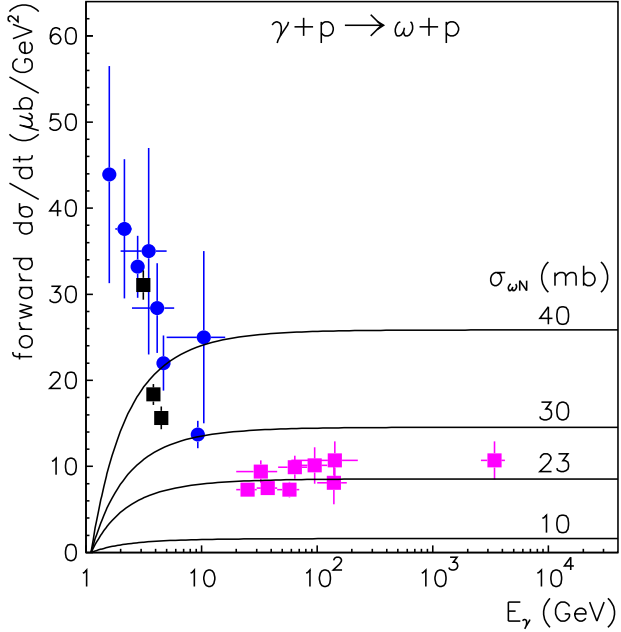,width=9.5cm,height=9.4cm}}
\vspace*{-2mm}
\caption{The forward $\gamma{p}{\to}\omega{p}$ differential cross
section as a function of photon energy. The data are collected in 
Ref.\cite{Sibirtsev1}.
The lines show the calculations based on Eq.~(\ref{vdm1}) 
with $\alpha_\omega$=0 and for different values of $\sigma_{\omega{N}}$.}
\label{spring3}
\end{figure}

Fig.~\ref{spring2a} demonstrates the
disagreement between the most recent measurements from SAPHIR~\cite{Barth}
and SPRING-8~\cite{Mibe}, which are shown by the solid circles and the solid
triangles, respectively. Note that both sets of data
contradict the VDM predictions. Although this discrepancy requires
special investigation we would like to make the following comments
relevant to the present study and VDM analysis of the data. 

\subsection{Comments}
Let us discuss the observed discrepancy between the data and the VDM 
description at low energies through inspection of
the forward $\gamma{p}{\to}\omega{p}$ differential cross section, 
shown in Fig.~\ref{spring3} as a function of the photon
energy. The different symbols indicate the data collected in
Ref.\cite{Sibirtsev3}, while the solid lines show the VMD calculations
based on Eq.~(\ref{vdm1}) with the ratio $\alpha_\omega = 0$ 
and for different values of
$\sigma_{\omega{N}}$, respectively. The data at $E_\gamma{>}$10~GeV
are in good agreement with the VDM assuming that
20${<}\sigma_{\omega{N}}{<}$30~mb. The low energy $\gamma{p}{\to}\omega{p}$ 
data indicate an excess with respect to VDM that is well understood in
terms of the non-diagonal $\rho{N}{\to}\omega{N}$ transition given by
Eq.~(\ref{ampli1}). More precisely, this can be explained by a $\pi$-meson
exchange contribution~\cite{Bauer,Friman}. Indeed,  the dominant 
$\omega$-meson decay mode is $\pi^+\pi^-\pi^0$, which in general 
is described in terms ofthe transition $\omega{\to}\rho\pi$, followed by the
$\rho{\to}\pi\pi$ decay. Therefore it is natural to expect that the
non-diagonal $\rho{N}{\to}\omega{N}$ transition plays a substantial
role in (low-energy) $\omega$-meson photoproduction. 

Most of the results on the forward photoproduction cross
section are determined through an extrapolation of the differential
$d\sigma{/}dt$ cross section over a certain range of $t$, not to 
the maximal accesible value, 
but rather to the point $t = 0$,  by applying a fit 
of the form  $d\sigma{/}dt{=}A\exp(bt)$. This is the reason why the $t =0$
differential cross sections at low photon energies, shown in
Fig.~\ref{spring3},  do not vanish even though they are clearly
dominated by $\pi$-meson exchange. The extrapolation to different $t$
might explain the difference between the SAPHIR~\cite{Barth} and
SPRING-8~\cite{Mibe} the results and data available at higher energies.

Indeed, within the Born approximation the $\pi$-meson exchange contribution 
vanishes~\cite{Sibirtsev3,Friman,Berman,Sibirtsev4}
at $t{=}0$. Here the minimal and maximal value of $t$ is given by
\begin{eqnarray}
t^\pm=m_V^2-\frac{s{-}m_N^2}{2s}\, \biggl(s+m_V^2-m_N^2
\nonumber \\ \mp\left[(s-m_V^2-m_N^2)^2
-4m_V^2m_N^2\right]^{1/2}\biggr)~,
\end{eqnarray} 
with $m_V$ and $m_N$ the masses of the vector meson and nucleon,
respectively. At threshold, $\sqrt{s}{=}m_V{+}m_N$, 
the four--momentum transfer squared is 
\begin{eqnarray}
t^\pm=-\frac{m_N \, m_V^2}{m_N+m_V},
\end{eqnarray}
With increasing energy, $\sqrt{s}{\gg}m_V{+}m_N$,
\begin{eqnarray}
t^-\simeq m_V^2-\frac{(s-m_N^2)\,  m_V^2 }{s},
\end{eqnarray}
and $t^-$ approaches zero. 

It is not obvious whether the extrapolation should be done to
$t{=}0$ or to $\theta{=}0$, as is explicitly shown by
Eq.~(\ref{optic}). Note that for elastic scattering $t^-{=}0$.
Actually the differential $\gamma{p}{\to}\phi{p}$
cross sections measured by SAPHIR~\cite{Barth} and SPRING-8~\cite{Mibe}
shown in Fig.\ref{spring2a} were extrapolated to $t{=}t^-$ and the
$t{=}0$ correction in that case is $\exp{(-t^-)}$,
which accounts for a factor of $\simeq 1.61$ at the $\phi$-meson
photoproduction threshold. 

Keeping that in mind one might conclude
that similar to the $\gamma{p}{\to}\omega{p}$ data the forward 
$\gamma{p}{\to}\phi{p}$ differential cross section indicates some
enhancement with respect to the diagonal VDM. This enhancement might 
stem from non-diagonal
transitions\footnote{Both SAPHIR and SPRING-8 measurements of angular
spectra in the Gottfried-Jackson frame support this conclusion.}.
Apparently that problem requires additional investigation, which is
beyond the scope of the current study.

Moreover the discrepancy between the  SAPHIR~\cite{Barth} and
SPRING-8~\cite{Mibe} results can be partially explained by the
different range of $t$ used for the $t{=}t^-$ extrapolation.
SPRING-8 explored the range $t{-}t^-{>} 0.4~{\rm GeV}^2$, while the
SAPHIR measurements were used to fit $d\sigma{/}dt$ over a
larger range and at maximum photon energy $t{-}t^- > -2~{\rm GeV}^2$.
For that reason it is worthwhile to
reanalyze the SAPHIR data by fitting the differential cross section with
the sum of a soft and a hard component.

Finally,  keeping in mind the uncertainty of the analysis of  low
energy data on the forward $\gamma{p}{\to}\phi{p}$ differential cross
section we conclude that VDM yields the upper limit of the total
$\phi{p}$ cross section about 11~mb. If the ratio of the
real to imaginary forward scattering amplitude, {\it i.e.}
$\alpha_\phi$, is not equal to zero, then $\sigma_{\phi{p}}$ can be
even smaller, as  indicated by Eq.(\ref{vdm1}).

\section{Single channel optical model}
Consider a nuclear reaction as a succession of collisions of the
incident particle with individual nucleons of the target. If the
nucleus is sufficiently large the reaction can be formulated in terms
of  the optical model through replacement of the multiple individual  
interactions by an effective potential interaction with nuclear matter. 
Within the so-called $t\rho$ approximation that is valid at normal nuclear
densities, $\rho$=0.16~fm$^{-3}$, the optical potential is given by the
product of the density and forward two-body scattering amplitude. Similarly,
the Glauber theory expresses the cross section for reactions on nuclear
target in terms of elementary two body
interactions~\cite{Glauber,Yennie,Bochmann1}. 

Conversely, by measuring
the nuclear cross sections it might be possible to study the
elementary interactions~\cite{Bauer}.  The realization of Drell and
Trefil~\cite{Drell} and Margolis~\cite{Margolis} that such a formalism
could be used to study the interaction of unstable
particles with the nucleon by producing them in a nucleus with hadronic or
electromagnetic beams, stimulated enormous experimental
activity~\cite{Bauer}. This method was first
applied~\cite{Drell,Margolis,Kolbig} for the evaluation of the $\rho{N}$
interaction from coherent and incoherent $\rho$-meson photoproduction
off nuclei. Coherent photoproduction has the advantage that the
produced particle must have quantum numbers similar to those of the
photon. 

\subsection{Coherent photoproduction}
An extensive study of  coherent $\phi$-meson photoproduction
on a variety of nuclear targets was
done at the Cornell 10 GeV electron
synchrotron~\cite{McClellan1}. One might expect~\cite{Margolis,Kolbig} the 
$A$--dependence of coherent photoproduction at high energies
to be proportional to $A^2$
if  the $\phi$-meson  does not interact in nuclei, 
{\it i.e.} if it is not distorted by  final state interactions (FSI).
By fitting the forward $\phi$ meson photoproduction cross
section~\cite{McClellan1} with a function $\sim A^\alpha$, one obtains
the slope $\alpha{=}1.37{\pm}0.08$ at the photon energy of 6.4~GeV and  
$\alpha{=}1.53{\pm}0.05$ at $E_\gamma = 8.3$~GeV.
The shaded boxes in Fig.~\ref{spring1b} show these results, 
which indicate a strong deviation from a dependence on $A^2$. 

Note, however, that the forward photoproduction cross
section contains both coherent and incoherent contributions. In
the absence of the FSI distortion, the incoherent photoproduction
cross section is proportional to $A$. Therefore one might argue that
the $A$-dependence  results from a
mixture of  coherent and incoherent $\phi$-meson photoproduction.
In order to verify such a possibility, an additional experiment with
linearly polarized photons with average energy of 5.7~GeV was
performed~\cite{McClellan2}. The measured polarization asymmetry 
for the forward
$\phi$-meson photoproduction from a carbon target was found to be consistent
with the assumption of coherent photoproduction\footnote{At the same time the
asymmetry from the hydrogen target seems to be inconsistent with purely
elastic $\phi$-meson photoproduction. As we discussed in Sec.~2
this Cornell observation~\cite{McClellan2} is in agreement with the most
recent results from SAPHIR~\cite{Barth} and SPRING-8~\cite{Mibe}.}.

\begin{figure}[t]
\vspace*{-6mm}
\centerline{\hspace*{3mm}\psfig{file=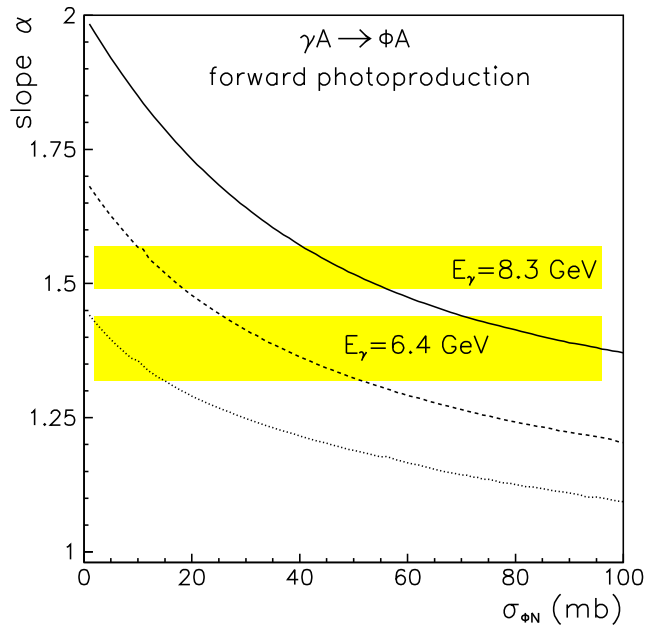,width=9.5cm,height=9.4cm}}
\vspace*{-3mm}
\caption{The slope $\alpha$ of the
$A^\alpha$ dependence of the forward $\phi$-meson photoproduction
cross section as a function of the total $\phi{N}$ cross section. The
shaded boxes indicate the results 
at photon energies 6.4 and 8.3 GeV obtained at
Cornell~\cite{McClellan1}.  The
lines show the calculations by Eq.(\ref{finalc}) with different
longitudinal momenta $q_l$=0 (solid), 63 (dashed) and 82~MeV/c
(dotted), which are fixed to the photon energy by Eq.(\ref{long}). The
calculations were done for the ratio $\alpha_\phi{=}0$.}
\label{spring1b}
\end{figure}

In order to evaluate the $\phi{N}$ interaction from coherent
photoproduction one can apply the method proposed in 
Refs.~\cite{Margolis,Kolbig} and express the 
amplitude for coherent $\phi$-meson photoproduction from nuclei  
as~\cite{Bauer,Drell,Kolbig,Ross1,Trefil1,Trefil2} 
\begin{eqnarray}
{\cal T}^{coh}_{\gamma{A}{\to}\phi A}(t,A)= 
{\cal T}_{\gamma{N}{\to}\phi N}
\!\int\limits^\infty_0\!d^2b 
\, J_0(q_t\,b)  
\nonumber \\  \times
\!\!\int\limits^\infty_{-\infty}\!\!dz \,\rho(b,z)
  \exp{[iq_lz]}\times\left[1-i\chi_{\phi}(b,z)\right]^{A-1},
\label{ampl1}
\end{eqnarray}
where ${\cal T}_{\gamma{N}}$ is the elementary photoproduction amplitude
on a nucleon, an integration is performed over the impact parameter $b$, the
$z$ coordinate is along the beam direction and 
$\rho(r{=}\sqrt{b^2{+}z^2})$ 
is the nuclear density function normalized to the number of the
nucleons in the nucleus.
Here, $t$ is the four-momentum transferred to the nucleus and
${-}t = q_l^2 + q_t^2$, with $q_l$ and  $q_t$ being  the 
longitudinal and transverse component, respectively,
given by
\begin{equation}
q_l = k-\cos\theta\sqrt{k^2-m_\phi^2}, \hspace{5mm}
q_t = \sin{\theta}\sqrt{k^2-m_\phi^2},
\label{long}
\end{equation}
where $k$ is the photon momentum, $m_\phi$ is the
pole mass of the $\phi$-meson and $\theta$ is the emission angle of the
produced $\phi$-meson. In Eq.~(\ref{ampl1})
$J_0$ is the zero order Bessel function. The last term of
Eq.~(\ref{ampl1}) accounts for the distortion of the  $\phi$-meson
through an effective interaction with  $A{-}1$ nuclear nucleons and  
$\chi_{\phi}$ is the corresponding nuclear phase shift. 
Here we neglect the distortion of the photon. 
The phase shift $\chi_{\phi}$ can be well approximated 
within the impulse approximation by~\cite{Newton,Sibirtsev5}
\begin{equation}
\chi_{\phi}(b,z) =  -\frac{2\pi \, f_{\phi{N}}(p_\phi,\theta{=}0)}{p_\phi}
\int\limits^\infty_z\!\rho(b,y)\, dy,
\label{phase1}
\end{equation}
where $f_{\phi{N}}$ is the complex amplitude for the forward
$\phi{N}$ elastic scattering taken now in the rest--frame with respect to
the nucleus, {\it i.e.} in the laboratory system. Note that $p_\phi$ is
$\phi$-meson momentum in the laboratory frame.
The imaginary part of $f_{\phi{N}}$ amplitude is related 
through the optical theorem to the total cross section 
$\sigma_{\phi{N}}$ similar to Eq.(\ref{optic}), replacing $q_\phi$
by $p_\phi$.

By introducing the ratio of the real to imaginary part of the forward
$\phi{N}$ scattering amplitude, $\alpha_\phi$, the  cross section for 
coherent $\phi$-meson photoproduction from nuclei is finally given as
\begin{eqnarray}
\frac{d\sigma_{\gamma A{\to}\phi A}^{coh}}{dt}=
\frac{d\sigma_{\gamma{N}{\to}\phi N}}{dt}
\left|\,
\int\limits^\infty_0\!d^2b\,\,
J_0(q_tb)\!\int\limits^\infty_{-\infty}\!\!dz \,\rho(b,z)
\right. \nonumber \\ \left.\times \exp{[iq_lz]} \,
\exp\left[\frac{\sigma_{\phi{N}}\, (i\alpha_\phi{-}1)}{2}
\int\limits^\infty_z\!\rho(b,y)\, dy\right]\right|^2\!\!\!\!.\,\,\,\,\,
\,\,\,\,\,
\label{finalc}
\end{eqnarray}
The general features of the coherent photoproduction are as
follows.  The $t$-dependence is given by the elementary 
$\gamma{N}{\to}\phi{N}$ photoproduction amplitude as well as by the
nuclear form factor. The differential cross section has a diffractive 
structure due to the $J_0(q_tb)$ dependence. However, up to now this 
structure was not observed experimentally, since it is non--trivial to 
isolate coherent from incoherent photoproduction, which dominates at 
large $|t|$. The forward coherent $\gamma{A}{\to}\phi{A}$
photoproduction cross section 
might be used for the extraction of the elementary forward
$\gamma{N}{\to}\phi{N}$ cross section, which can be compared with
those collected in Fig.\ref{spring2}. The $A$--dependence of 
coherent photoproduction allows one to extract $\sigma_{\phi{N}}$ only
under certain constraints on $\alpha_\phi$. That extraction
is independent of the VDM assumptions. 

\begin{figure}[t]
\vspace*{-8mm}
\centerline{\hspace*{3mm}\psfig{file=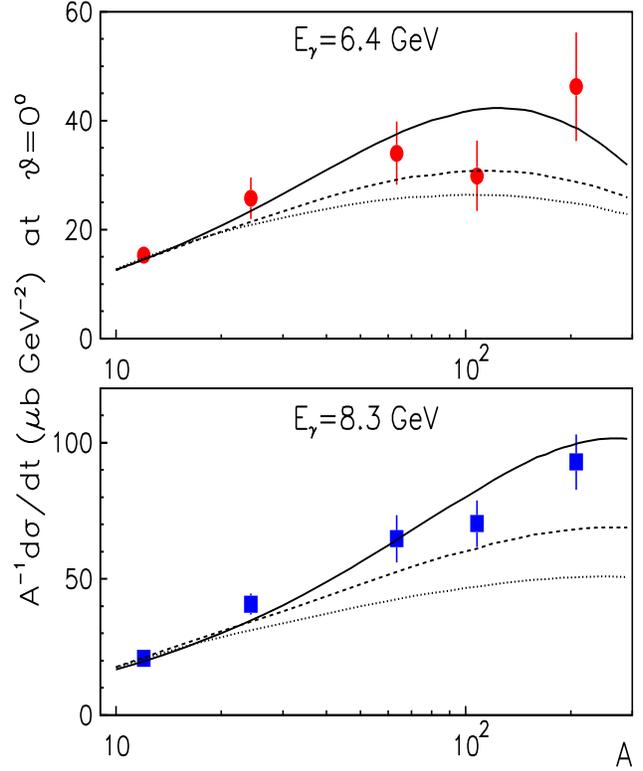,width=9.5cm,height=12.cm}}
\vspace*{-7mm}
\caption{The forward $\gamma{A}{\to}\phi{A}$ differential cross
section as a function of the mass number. The symbols show the data
collected at Cornell~\cite{McClellan1} at photon energies 6.4
(circles) and 8.3~GeV (squares). The lines are the calculations with
the total $\phi{N}$ cross section of 10 (solid), 30 (dashed) and 50~mb
(dotted) and with the ratio $\alpha_\phi{=}0$. Both experimental
results and calculations are divided by $A$. The calculations are
normalized at the $A{=}12$ point.}
\label{spring5b}
\end{figure}

Indeed if the distortion of the $\phi$-meson is negligible and
$q_l{=}0$ the coherent photoproduction cross section is proportional
to $A^2$ as is given by Eq.(\ref{finalc}). Furthermore, to analyze the Cornell
data and for completeness, we specified the density distribution
function $\rho(r)$ used in Eq.(\ref{finalc}) as
\begin{eqnarray}
\rho(r)=\frac{\rho_0}{1+\exp[(r-R)/d]},
\end{eqnarray}
with parameters
\begin{eqnarray}
R{=}1.28A^{1/3}{-}0.76{+}0.8A^{-1/3}\,\, {\rm fm},
\,\,\,\,\,\, d=\sqrt{3}/\pi \,\, {\rm fm},
\end{eqnarray}
for the nuclei with $A{>}16$. For light nuclei we
adopt~\cite{Dalkarov,Alkhasov} 
\begin{eqnarray}
\rho(r)=(R\sqrt{\pi})^{-3}\left[4+\frac{2(A-4)r^2}{3R^2}
\right]\exp[-r^2/R^2],
\end{eqnarray}
with $R{=}\sqrt{2.5}$~fm.

We note that the results on forward photoproduction are not
sensitive to the variation of the ratio $\alpha_\phi$, as can be seen
from Eq.(\ref{finalc}) and is known from the analyses of $\rho$-meson
photoproduction~\cite{Bauer} and hadronic elastic
scattering~\cite{Sibirtsev5}. 

Finally, the lines in Fig.~\ref{spring1b} are
the calculations based on Eq.~(\ref{finalc}) with momentum $q_l = 82$ 
(dashed) and 63 (dotted) and 0~MeV/c (solid),  related to the
photon energy by Eq.~(\ref{long}). The results are shown as a function
of the total $\phi{N}$ cross section. The calculations for  $q_l$=0~MeV/c
correspond to the high energy limit, {\it i.e.} $k{\gg}m_\phi$, and 
for $\sigma_{\phi{N}}$=0 actually match the $A^2$
point. To get the slope $\alpha$ we fit our calculation with the function
$cA^\alpha$ with $c$ a constant and using the set of nuclear
targets corresponding to the experimentals.
Fig.~\ref{spring1b} illustrates that the $q_l$ correction
already introduces a substantial departure from the
$A^2$-dependence. Although the uncertainties in the  $\sigma_{\phi{N}}$
extraction from the data are large, the data are in very good
agreement with the calculations with a total $\phi{N}$ cross section 
of $\simeq 10$~mb.

Fig.~\ref{spring5b} shows the forward $\gamma{A}{\to}\phi{A}$
differential cross section as a function of the target mass 
number $A$. The circles and
squares are the experimental results obtained at Cornell for photon
energies of 6.4 and 8.3~GeV, respectively. The lines show the
calculations for different $\sigma_{\phi{N}}$ and for the ratio   
$\alpha_\phi{=}0$. Both data and calculations are 
divided\footnote{Although in view of Eq.(\ref{finalc}) it is more natural
to  divide the results on coherent photoproduction by $A^2$, it turns
out that $A^{-1}$ representation is more illustrative in case of the observed
moderate $A$-dependence of the data and therefore is
very frequently used.} by $A$. Obviously the shape of the
$A$-dependence is different for the calculation  for the various
$\sigma_{\phi{N}}$. The experimental results are in perfect agreement with
the calculations using $\sigma_{\phi{N}}{=}10$~mb.

Finally,  one can as well extract the
elementary $\gamma{N}{\to}\phi{N}$ forward differential cross section
using Eq.(\ref{finalc}) and compare the results with the results
obtained by direct measurement. 
The calculations with $\sigma_{\phi{N}}$=10 mb can be well fitted to
the data with an elementary $\gamma{N}{\to}\phi{N}$ forward differential
cross section around $2.2{-}2.6$ $\mu$b/GeV$^2$, which is in good agreement
with the results collected in Fig.\ref{spring2}.

\subsection{Incoherent photoproduction}
Recently incoherent $\phi$-meson photoproduction at photon energies
1.5${\le}E_\gamma{\le}$2.4~GeV was studied by the SPRING-8
Collaboration~\cite{Ishikawa}. The data were published with arbitrary
normalization and provide the $A$ dependence fitted by the function
$A^\alpha$ with slope $\alpha{=}0.72{\pm}0.07$.

The optical model expression for the incoherent photoproduction
including only the excitation of the single nucleon and
neglecting the Pauli principle, which suppresses the cross section at small
$t$, is given by~\cite{Bauer}
\begin{eqnarray}
\frac{d\sigma_{\gamma A{\to}\phi A}^{inc}}{dt}{=}
\frac{d\sigma_{\gamma{N}{\to}\phi N}}{dt}\!\!
\int\limits^\infty_0\!\!d^2b\!\!
\!\int\limits^\infty_{-\infty}\!\!\!dz \,\rho(b,z) \nonumber \\
\exp\Bigl[{-}\sigma_{\phi{N}}\!\!\!\int\limits^\infty_z\!\!\!\rho(b,y)\,
dy\Bigr]~.
\label{inco1}
\end{eqnarray}
Now if $\sigma_{\phi{N}} = 0$, the forward cross section is 
proportional to $A$. The
shaded box in Fig.~\ref{spring1} shows the result from SPRING-8, while
the line indicates the calculations based on Eq.~(\ref{inco1}) for different
values of the total $\phi{N}$ cross section. Again, we fit our results
by the function $cA^\alpha$ and use the set of nuclear targets from
experiment\footnote{The $A^\alpha$ function is not the  dependence
given by  Eq.~(\ref{inco1}) and  only provides a useful representation of
the data. Indeed the slope $\alpha$ depends on the set of target
numbers $A$ used in the calculations. Therefore it is necessary to
simulate the experimental conditions explicitely.}. It is clear that the
experimental results favor $ 23 \le \sigma_{\phi{N}} \le  63$~mb. This is in
agreement with the experimental finding~\cite{Ishikawa} given as
$\sigma_{\phi{N}}= 35^{+17}_{-11}$~mb. The Pauli blocking corrections
make almost no change to the $A$-dependence and only suppress the absolute
value of the forward photoproduction cross section~\cite{Bauer,Cabrera}.

\begin{figure}[t]
\vspace*{-6mm}
\centerline{\hspace*{3mm}\psfig{file=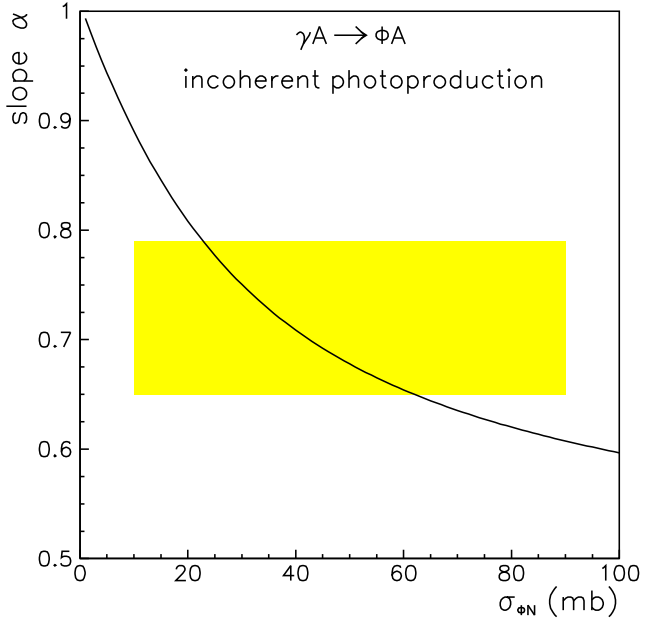,width=9.5cm,height=9.4cm}}
\vspace*{-4mm}
\caption{The slope $\alpha$ of the
$A^\alpha$ dependence of the incoherent $\phi$-meson photoproduction
cross section as a function of the total $\phi{N}$ cross section. The
shaded boxes indicate the results 
at photon energies 1.5${\le}E_\gamma{\le}$2.4~GeV obtained by
SPRING-8~\cite{Ishikawa}.  The
lines show the calculations by Eq.(\ref{inco1}).}
\label{spring1}
\end{figure}

Clearly this result  differs substantially from the total
$\phi{N}$ cross section extracted from coherent $\phi$-meson
photoproduction. There are no available explanations why the distortion
of the incoherently produced $\phi$-meson is so extremely strong.
Furthermore, the calculations~\cite{Cabrera,Munlich} wich include  the
in-medium  modification of the $\phi$-mesoncannot account for such strong
distortion. 

\section{Coupled channel scattering}
Since $\phi$ and $\omega$ mesons have the same quantum numbers, these
two states should mix with each other - see e.g.~\cite{Ross2,Trefil3,Sandra}. 
Therefore the $\phi$-meson might be produced indirectly, 
{\it i.e.} through the photoproduction of the $\omega$-meson followed
by  the $\omega{\to}\phi$ transition. Furthermore, there can in
principle occur an arbitrary
number of  $\omega{\leftrightarrow}\phi$ transitions. Moreover,
these transitions can occur either due to the $\omega$ and $\phi$
mixing, {\it i.e.} similar to oscillations, or because of the
$\omega{N}{\to}\phi{N}$ interaction. This is illustrated in Fig.~\ref{fig:mix}.
Since both $\omega$ and $\phi$
mesons are strongly interacting particles, their final state interaction
in any nuclear target would cause distortions. 

\begin{figure}[t]
\vspace*{2mm}
\centerline{\hspace*{3mm}\psfig{file=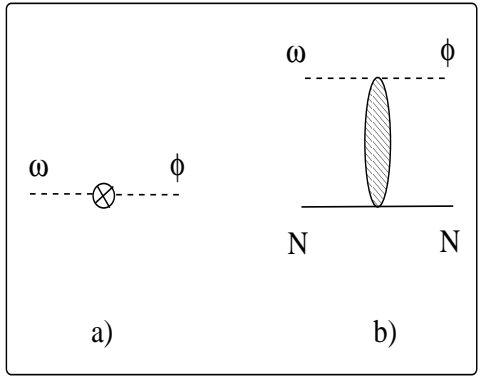,width=6.3cm,height=5.cm}}
\caption{Mechanisms of $\omega \to \phi$ conversion: a) direct 
$\omega-\phi$ mixing amplitude and b) the $\omega{N}{\to}\phi{N}$ interaction,
where $N$ denotes a nucleon. Type b)  contains the direct mechanism a) besides 
other effects. } 
\label{fig:mix}
\end{figure}

The coupled channel scattering
apparently depends on the strength of the
$\omega{\leftrightarrow}\phi$ transition. However, one 
expects the $\phi$-meson distortion to be considerable because of
the interference effect. Even a small amount of an
$\omega{-}\phi$ admixture might have a considerable effect on
$\phi$-meson production from complex nuclei. First discussed by Ross
and Stodolsky~\cite{Ross1,Ross2}, this effect has subsequently been largely
overlooked in the literature~\cite{Bauer}. Here we consider coherent
and incoherent $\phi$-meson photoproduction from nuclei within the
coupled channel approach.

\subsection{Coherent photoproduction}
The generalization of the single channel amplitude of
for the two coupled channel scattering can be done 
by introducing the 2$\times$2 matrix instead of the last term of
Eq.(\ref{ampl1}) as
\begin{eqnarray}
\left( \begin{array}{cc}
\!\!\!\!\!\!\!\!1{-}\displaystyle\frac{\sigma_{\omega{N}}}{2} 
\int\limits^\infty_z\!\!\rho(b,y)\,dy
& \!\!\!\!\!\!\!\!\!\!\!\displaystyle\frac{{-}\Sigma_{\omega\phi}}{2}
\!\!\!
\int\limits^\infty_z\!\!
\exp[i{\tilde q}_ly]\rho(b,y)dy\\
 & \\
\displaystyle\frac{{-}\Sigma_{\omega\phi}}{2}\!\!\!
\int\limits^\infty_z\!\!
\exp[-i{\tilde q}_ly]\rho(b,y)dy & \!\!\!\!
1{-}\displaystyle\frac{\sigma_{\phi{N}}}{2}
\!\!\! 
\int\limits^\infty_z\!\!\rho(b,y)\,dy
\end{array} \right)\,\,\,
\label{matrix}
\end{eqnarray}
where we have used Eqs.~(\ref{optic},\ref{phase1}) and we denote by $\Sigma$
the effective $\omega{\to}\phi$ and  $\phi{\to}\omega$ transition 
cross section, which are equivalent because of time-reversal
invariance. Furthermore, we neglect the real parts of the elastic and
transition amplitudes, since even in the single channel analysis they
could not be fixed. Furthermore, the longitudinal momentum 
${\tilde q}_l$ is defined as
\begin{eqnarray}
{\tilde q}_l \simeq \frac{m_\phi^2-m_\omega^2}{2k}~.
\end{eqnarray}
Neglecting the off-diagonal transition, {\it i.e.} setting $\Sigma = 0$, 
the matrix of Eq.~(\ref{matrix}) allows one to recover the single channel
optical model for $\omega$ and $\phi$-meson photoproduction, while
taking the elementary ${\cal T}_{\gamma{N}}$ amplitude as a two-component
vector.

\begin{figure}[t]
\vspace*{-8mm}
\centerline{\hspace*{3mm}\psfig{file=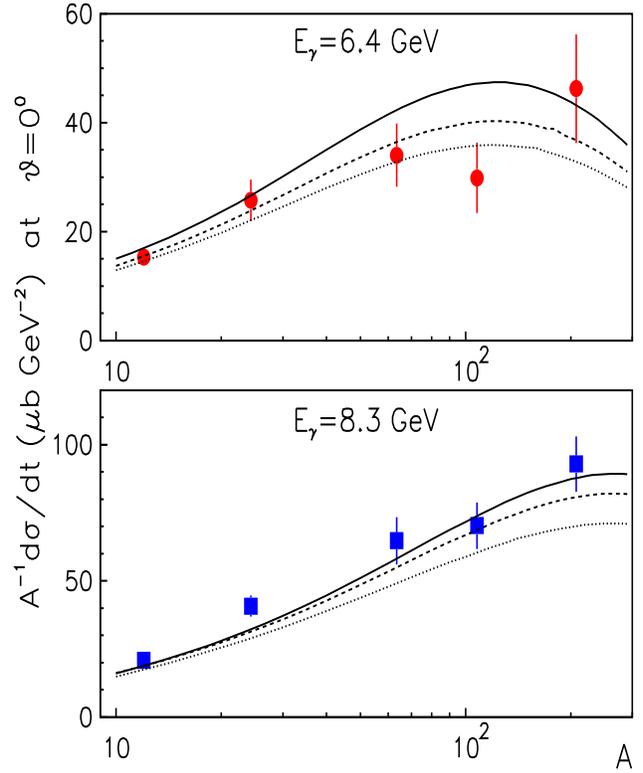,width=9.5cm,height=12.cm}}
\vspace*{-7mm}
\caption{The forward $\gamma{A}{\to}\phi{A}$ differential cross
section as a function of the mass number. The symbols show the data
collected at Cornell~\cite{McClellan1} at photon energies 6.4
(circles) and 8.3~GeV (squares). The lines are the coupled channel scattering
calculations by Eq.(\ref{two}) with the total $\phi{N}$ cross section
of  11~mb and the $\omega{N}$ cross section of 23~mb and for the
transition $\Sigma$=0 (solid), 0.3 (dashed) and 0.5~mb (dotted).
 Both experimental results and calculations are divided by $A$. The
normalization of the calculations is fixed by VDM as explained in the
text.}
\label{spring5c}
\end{figure}

Since in our case  $\Sigma$ is small, the amplitude for 
coherent $\phi$-meson photoproduction from a nucleus can be expressed in
a simple form, 
\begin{eqnarray}
{\cal T}_{\gamma A{\to}\phi A}^{coh}{=}\int\limits^\infty_0\!\!
d^2b\,\,  J_0(q_tb) \!\! 
\int\limits^\infty_{-\infty}\!\! dz \, \rho(b,z) \, \exp[iq_lz] 
\nonumber \\ \times\left(  {\cal T}_{\gamma N{\to}\omega N}
\displaystyle\frac{-\Sigma \int\limits^\infty_z\,\,  \exp[i{\tilde q}_ly]\,\,
\rho(b,y)\, dy}{(\sigma_{\phi N}{-}\sigma_{\omega N})
\int\limits_z^\infty \rho(b,y)\, dy}  \right. \nonumber \\ \left.
 \times \left[\exp[-\frac{\sigma_{\omega N}}{2}\!\!\!\int\limits_z^\infty
\!\!\rho(b,y)\, dy]{-}
\exp[-\frac{\sigma_{\phi N}}{2}\!\!\!\int\limits_z^\infty
\!\!\rho(b,y)\,dy]\right] \nonumber \right. \\ \left.
+{\cal T}_{\gamma N{\to}\phi N}\exp[-\frac{\sigma_{\phi N}}{2}
\!\!\!\int\limits_z^\infty\!\!\rho(b,y)\,dy]\right).\,\,\,\,
\label{two}
\end{eqnarray}
For $\Sigma{=}0$, Eq.~(\ref{two}) reduces to Eq.~(\ref{ampl1}) and
the coherent differential cross section is given by
Eq.~(\ref{finalc}). Moreover, Eq.~(\ref{two}) corresponds to the first
order perturbation expansion in $\Sigma$, {\it i.e.} the inclusion of only
one $\omega{\to}\phi$ transition. If $\Sigma$ is large one should
consider an arbitrary number of  $\omega{\leftrightarrow}\phi$
transitions, which might be done through the series
expansion~\cite{Benhar} of the off--diagonal elements of the matrix given
by Eq.~(\ref{matrix}).

Moreover in high energy limit, {\it i.e.} when $q_l{=}{\tilde
q}_l = 0$, the integration along $z$ can be done analytically and the
coherent  photoproduction amplitude is then  given in a well-known 
form~\cite{Kolbig}
\begin{eqnarray}
{\cal T}_{\gamma A{\to}\phi A}^{coh}{=}\int\limits^\infty_0\!\!
d^2b\,\,  J_0(q_tb) \left[ \frac{{\cal T}_{\gamma N{\to}\omega
N}\,\,\Sigma}
{\sigma_{\omega{N}}{-}\sigma_{\phi{N}}} \right. \,\,\,\,\nonumber \\
\times\left(\frac{1{-}\exp{[-\displaystyle\frac{\sigma_{\omega{N}} \, T(b) }{2}
]}}{\sigma_{\omega{N}}}
-\frac{1{-}\exp{[-\displaystyle\frac{\sigma_{\phi{N}} \, T(b)}{2}]}}
{\sigma_{\phi{N}}}\right) \nonumber
\\\left.+\frac{{\cal T}_{\gamma N{\to}\phi
N}}{\sigma_{\phi{N}}}
\left(1{-}\exp{[-\displaystyle\frac{\sigma_{\phi{N}} \, T(b)}{2}]}
\right)\right],\,\,\,\,\,\,
\label{limit}
\end{eqnarray}
where the real parts of the scattering amplitudes were neglected and
the  thickness function is given as
\begin{eqnarray}
T(b)=\!\!\int\limits_{-\infty}^\infty
\!\!\rho(b,z)\,dz.
\label{tb}
\end{eqnarray}

Following the VDM results shown in Figs.\ref{spring2},\ref{spring3} we
use the  elementary elastic $\omega$ and $\phi$-meson photoproduction
amplitude given by Eq.\ref{vdm1}, namely as 
\begin{eqnarray}
{\cal T}_{\gamma
N{\to}VN}=\frac{\sqrt{\alpha}}{4\gamma_V}\frac{q_V}{q_\gamma}
\sigma_{VN},
\end{eqnarray}
with the $\gamma_\phi$ and $\gamma_\omega$ coupling constants given by
Eq.(\ref{coupl2}) and with
$\sigma_{\phi N}{=}11$~mb and $\sigma_{\omega N}$=23~mb. Thus the
absolute normalization of our calculations is fixed by VDM. 
Fig.\ref{spring5c} shows the coherent  $\gamma{A}{\to}\phi{A}$
differential cross section as a function of the mass number calculated
for different values of $\Sigma$. The data~\cite{McClellan1} might be
well reproduced by calculations with
$0{\le}\Sigma{\le}0.3$~mb. Only the experimental results at
$E_\gamma{=}6.4$~GeV support the coupled channel effect originating
from the $\omega{\to}\phi$ transition. 

\subsection{Incoherent photoproduction}
Let us now consider incoherent photoproduction of $\omega$-mesons followed by
$\omega{N}{\to}\phi{N}$ scattering. Note that in the coupled
channel description of 
coherent $\phi$-meson photoproduction the $\omega{\to}\phi$
transition is not necessaryly due to  the scattering on the target nucleon, but
might be also an oscillation due to the mixing. In that sense
incoherent photoproduction is given as a two step process and
differential cross section can be written 
as~\cite{Bochmann1,Faldt,Vercellin}
\begin{eqnarray}
\frac{d\sigma_{\gamma A{\to}\phi A}^{inc}}{dt}{=}
\frac{d\sigma_{\gamma{N}{\to}\omega N}}{dt}\!\!
\int\limits^\infty_0\!\!{\tilde \Sigma} \, d^2b
\left[\frac{1-\exp[-\sigma_{\phi N}\, T(b)]}{\sigma_{\phi N}}
\right. \nonumber \\
\left. -\frac{\exp[-\sigma_{\omega N}\, T(b)]{-}\exp[-\sigma_{\phi N}\,
T(b)]}
{\sigma_{\phi N}-\sigma_{\omega N}} \right],\,\,\,\,\,
\label{inco2}
\end{eqnarray}
where the function $T(b)$ is given by Eq.(\ref{tb}) and it is not necessary
that $\Sigma = {\tilde \Sigma}$. Considering both  direct
and two-step $\phi$-meson production, one should in principle
add the contribution  given by Eq.(\ref{inco1}). 

Unfortunately the data~\cite{Ishikawa} on incoherent $\phi$-meson 
photoproduction from nuclei are given with arbitrary normalization and
we cannot investigate how big the possible contribution from
two-step production might be, since ${\tilde\Sigma}$ is unknown. However, it is
possible to examine the $A$-dependence due to the two-step process.   

\begin{figure}[t]
\vspace*{-6mm}
\centerline{\hspace*{3mm}\psfig{file=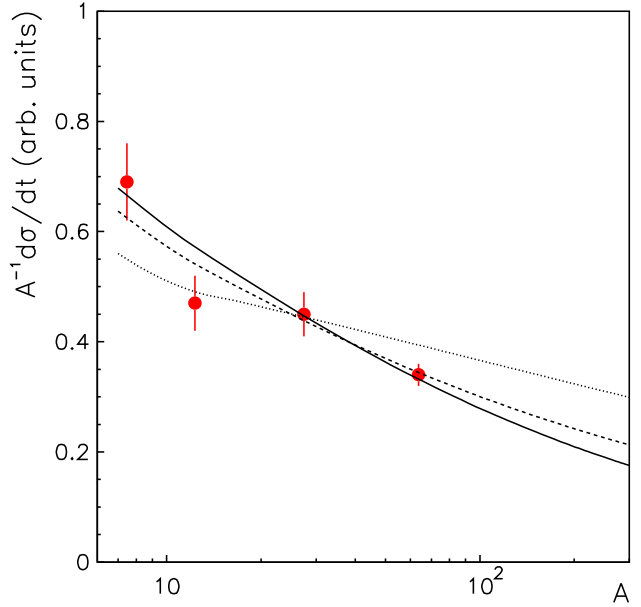,width=9.5cm,height=9.4cm}}
\vspace*{-7mm}
\caption{The incoherent $\phi$-meson photoproduction  cross
section as a function of the mass number. The circles show the data
collected at SPRING-8~\cite{Ishikawa}. The solid line is the coupled 
channel scattering calculations by Eq.(\ref{inco2}) with the total 
$\phi{N}$ cross section of  11~mb and the $\omega{N}$ cross section of
23~mb, while the dashed line is result obtained with   $\sigma_{\phi
N}$=11~mb and  $\sigma_{\omega N}$=30~mb.
The dotted line is the single channel results for $\sigma_{\phi N}$=11~mb.
Both experimental results and calculations are divided by $A$. The
normalization of the calculations is fixed at the $Al$ target.}
\label{spring6}
\end{figure}

Fig.~\ref{spring6} shows the incoherent $\phi$-meson photoproduction
cross section as a function of the target mass
measured at SPRING-8~\cite{Ishikawa}. The dotted line shows the
calculations performed within the single channel optical model using
Eq.(\ref{inco1}) with $\sigma_{\phi N} = 11$~mb.
The solid line represents the result obtained with the
coupled channel model calculation, Eq.(\ref{inco2}),
using $\sigma_{\phi N} = 11$~mb and $\sigma_{\omega N} = 23$~mb. 
Both the data and our results are divided by $A$. The calculations are 
normalized for an $Al$ target. The single channel calculations are 
identical to the ones shown in
Refs.~\cite{Ishikawa,Cabrera,Munlich} and apparently can not
reproduce the data, as we already discussed in Sec.3.2. The two-step 
model calculations are in reasonable agreement with experimental
results - providing an $A$-dependence ${\propto}A^{0.63}$. Thus the
measured $A$-dependence clearly indicates the dominance of the two-step
process in $\phi$-meson photoproduction. Unfortunately there are no
data available for heavy targets, which clearly are crucial for the
verification of the calculated $A$-dependence shown in
Fig.~\ref{spring6}.  

Note that another two-step process, {\it i.e.} $\gamma{N}{\to}\pi{N}$
followed by $\pi{N}{\to}\phi{N}$ transition, would produce almost the same
dependence, ${\propto}A^{0.63}$, as shown by the dashed line in
Fig.\ref{spring6}. In the considered $\pi$-meson momentum range the
total $\pi{N}$ cross section is about $\simeq$30~mb and therefore
calculations were done using Eq.(\ref{inco2}) with $\sigma_{\omega
N}$=30~mb. Note that both $\pi$-meson photoproduction and the
$\pi{N}{\to}\phi{N}$ transition are quite large and therefore
it is possible that  incoherent $\phi$-meson photoproduction is
dominated by a two--step reaction mechanism. For instance theoretical 
studies~\cite{Cassing1,Sibirtsev6,Debowski,Sibirtsev7,Paryev} on low
energy $K^+$, $\rho$, $\omega$ and $\phi$-meson production in proton
nucleus collisions indicate dominance of the two-step process with
intermediate $\pi$-mesons. This result is strongly supported by
measurements of the $A$-dependence and two-particle correlations in
$K^+$-meson production from $pA$
reactions~\cite{Debowski,Koptev1,Badala,Buescher1,Koptev2}. 
Further investigations on incoherent $\phi$-meson photoproduction
require measurement of  differential cross section with
absolute normalization.

It is also clear that incoherent coupled channel $\phi$-meson
photoproduction might also proceed through  coherent $\omega$
meson production by the photon, followed by the incoherent
$\omega{\to}\phi$ transition. This mechanism involves, in addition,
substantial $q_l$-dependence at low photon energies 
similar to that of Eq.(\ref{finalc}), which allows
for  freedom in description of the $A$-dependence. 

\subsection{Estimates for {\boldmath$\Sigma$} and 
{\boldmath${\tilde\Sigma}$}}

It is useful to estimate the $\Sigma$ and ${\tilde\Sigma}$ 
in order to understand how large the effect due to the
$\omega{\to}\phi$ transition might be. Our
estimates are based on the amplitudes evaluated in free space, which 
are not necessarily the same as in nuclear matter. 

Very recently the the $\omega{-}\phi$ mixing amplitude
$\Theta_{\omega\phi}$  was investigated~\cite{Kucurkarslan} within 
the leading order chiral perturbation theory and it was found that
$\Theta_{\omega\phi}{=}(25.34{\pm}2.39){\times}10^{-3}$~GeV$^2$. In our
normalization $\Sigma$ can be related to $\Theta_{\omega\phi}$ 
as~\cite{Urech}
\begin{eqnarray}
\Sigma \simeq \frac{1}{m_\omega^2}
\frac{\Theta_{\omega\phi}^2}{(m_\phi^2-m_\omega^2)^2} = 2.2
{\rm \mu b},
\end{eqnarray}
where we neglect the width of the $\omega$ and $\phi$-meson.
Actually the effect due to the $\omega{-}\phi$ mixing is small and as
is indicated by the calculations shown in Fig.\ref{spring5c} might be
supported by the Cornel data~\cite{McClellan1} on coherent
$\phi$-meson photoproduction from nuclei. However, the data itself do
not really require the inclusion of the mixing amplitude and the problem still
remains open. 

The contribution to incoherent $\phi$-meson photoproduction from the 
two step process with an intermediate
$\pi$-meson can be reasonably estimated since there are data available for
the $\pi{N}{\to}\phi{N}$ reaction collected in
Refs.\cite{Sibirtsev8,Sibirtsev9} and parameterized as
\begin{eqnarray}
{\tilde\Sigma}(\pi{N}{\to}\phi{N})=\frac{18\sqrt{s-s_0}}
{0.1285+(s-s_0)^2}\,\,\ {\rm \mu b},
\end{eqnarray}
where $s$ is the squared invariant mass of the $\pi{N}$ system given
in GeV$^2$ and
$\sqrt{s_0}{=}m_N{+}m_\phi$ is the reaction threshold. Note that at pion
energies of$\simeq 2$~GeV, which correspond to the forward $\phi$-meson
photoproduction in the SPRING-8 experiment~\cite{Ishikawa} the
$\pi{N}{\to}\phi{N}$ cross section is about 20~$\mu$b. Furthermore, at
photon energies of 1.5${\le}E_\gamma{\le}$2.4~GeV the total cross section
for the $\gamma{N}{\to}\phi{N}$ reaction is in average
$\simeq$0.3~$\mu$b~\cite{Sibirtsev10}
while the  $\gamma{N}{\to}\pi{N}$ reaction accounts for
$\simeq$5~$\mu$b~\cite{Kellett,Landolt}. Therefore the contribution
from the two-step mechanism might be well suppressed as compared to the
direct $\phi$-meson photoproduction. This can be clearly seen by
inspecting Eqs.(\ref{inco1},{\ref{inco2}) and replacing the $\omega{N}$
intermediate state with the $\pi{N}$ one. 

It is difficult to estimate reliably the contribution to incoherent
$\phi$-meson photoproduction from a two-step process
with an intermediate $\omega$-meson. Figs.\ref{spring2a},\ref{spring3}
illustrate that incoherent forward
$\omega$-meson photoproduction dominates
$\phi$-meson photoproduction  by a factor of order$\simeq 60$. 
However, at the same time the 
$\gamma{p}{\to}\phi{p}$ data collected in Fig.\ref{spring3}
show almost no room for the non-diagonal $\omega{N}{\to}\phi{N}$ 
transition. The data at low photon energies are well described by the 
VDM accounting only for elastic diagonal $\phi{N}{\to}\phi{N}$
scattering. Within the experimental uncertainties of the forward
$\gamma{p}{\to}\phi{p}$ differential cross sections and considering the
difference between data and our VDM calculations with $\sigma_{\phi
N}$=11~mb we estimate ${\tilde\Sigma}{<}$0.1~mb. In that case the
contribution from the two-step process is compatible with the direct
incoherent $\phi$-meson photoproduction and the coupled channel effect is
indeed sizable.  

\subsection{Speculations}
Finally we would like to mention another possibility which is not
related to the $\phi$-meson propagation in nuclear matter and the
${\tilde\Sigma}$ transition but with incoherent $\phi$-meson 
photoproduction at low energies. Recently~\cite{Sibirtsev11} 
we investigated the role of
the cryptoexotic baryon with hidden strangeness, $B_\phi{=}udds{\bar
s}$, in $\phi$-meson production in proton-proton collisions close to
the reaction threshold. We found that the enhanced $\phi$-meson production
observed at COSY~\cite{Hartmann1} can be well explained by
an $B_\phi$-baryon excitation followed by the  $B_\phi{\to}\phi{N}$
decay. It is expected that these
pentaquark baryons have a narrow width and decay preferentially into the
$\phi{N}$, $K{\bar K}N$ or $YK$ channels, where $Y$ stands for ground-state
or excited hyperons~\cite{Landsberg1,Landsberg2}.
Experimental observations for the $B_\phi$ candidates were reported
in Refs.~\cite{Arenton,Aleev,Dorofeev1,Dorofeev2,Balatz,Antipov}.
The high-statistics study of Ref.~\cite{Antipov} of the 
$\Sigma^0K^+$ mass spectrum indicates two exotic states with 
$M = 1807{\pm}7$~MeV, $\Gamma = 62{\pm}19$~MeV and $M = 1986{\pm}6$~MeV,
$\Gamma = 91{\pm}20$~MeV. 

One might expect that $B_\phi$-baryon can be excited in photon-nucleon
interaction. Because of its mass the $\gamma{p}{\to}\phi{p}$ reaction
would be sensitive to $B_\phi$ excitation at low photon energy. The
$B_\phi$ contribution might explain the SAPHIR-8 and SPRING-8
measurements~\cite{Barth,Mibe} of angular spectra in the
Gottfried-Jackson frame, which indicate that at low energies the
$\phi$-meson photoproduction is not governed by pomeron exchange.

Considering coherent and incoherent $\phi$-meson photoproduction from
nuclei we notice even more significant features. First, the coherent
photoproduction at low energies should not by dominated by
$B_\phi$-baryon excitation, since in that case the residual nucleus
differs from ground state. Incoherent $\phi$-meson
photoproduction at low energies might be dominated by $B_\phi$
excitation. Since the $B_\phi$-baryon is narrow, it decays outside the
nucleus and an effective distortion of the $\phi$-meson is given by the
distortion of $B_\phi$, which is compatible with and interaction of
other baryons in nuclear matter because of the light quark content of
the $B_\phi$-baryon. Therefore one should not be surprised by the result
shown in Fig.~\ref{spring1}. 

\section{Conclusions}
We have analyzed coherent and incoherent $\phi$-meson photoproduction from
nuclei by applying the single and coupled channel optical model. 

The data
on coherent $\phi$ photoproduction collected at Cornell~\cite{McClellan1} at 
photon energies of 6.4 and 8.3~GeV can be well reproduced by a single
channel calculation taking into account the $\phi$-meson distortion
compatible with the $\phi{N}$ total cross section $\sigma_{\phi
N} =10$~mb. This result is in good agreement with the VDM analysis of
the forward $\gamma{p}{\to}\phi{p}$ differential cross section, which
indicates that $\sigma_{\phi N}{\simeq}$11~mb. Coherent
$\phi$-meson photoproduction shows little room for the coupled channel
effect due to the contribution from the $\omega {\to} \phi$ transition. 

The data on incoherent $\phi$-meson photoproduction off various nuclei 
collected at 
SPRING-8~\cite{Ishikawa} at photon energies from 1.5 to 2.4~GeV can be
reproduced by the single channel optical model calculations only under
the assumption that the $\phi$-meson is substantially distorted in nuclei,
which corresponds to $23 {\le} \sigma_{\phi N} {\le} 63$~mb. This result is
in agreement with previous incoherent $\phi$-meson photoproduction data
analyses~\cite{Ishikawa,Cabrera,Munlich}. Moreover, we
found that taking into account the coupled channel effects, {\it i.e.}
assuming direct $\omega$-meson photoproduction followed by the
$\omega{N} {\to} \phi{N}$ transition as well as pion photoproduction
followed by the $\pi{N} {\to} \phi{N}$ scattering, it is possible to reproduce
the $A$-dependence measured at SPRING-8~\cite{Ishikawa}. 

Although we
estimate the absolute rates for the contribution of these two
different intermediate states, it is difficult to draw a  final
conclusion. First, the SPRING-8 data~\cite{Ishikawa} are published
without absolute normalization. Second, the $t$-dependence of these
data are not given for all nuclei used in the measurements. 
Such knowledge is essential for the evaluation of  incoherent
photoproduction within the coupled channel analysis, since each of the
two-step processes has an individual $t$-dependence, which can be used in
order to distinguish the intermediate states. In that sense more
precise data on incoherent $\phi$-meson photoproduction are necessary
for further progress.
 
We also discussed a very alternative (and probably speculative) 
scenario that might occur only in
incoherent $\phi$-meson photoproduction but is not accessible in
the coherent reaction. The excitation of the cryptoexotic baryon with
hidden strange{\-}ness, called $B_\phi$, would result in an
$A$-dependence similar to that measured by SPRING-8~\cite{Ishikawa}.
However, the $B_\phi$-baryon could not be exited in coherent
photoproduction. Therefore measurements of the $A$-dependence of 
coherent $\phi$-meson photoproduction from nuclei at low energies is
crucial for identification of the possible existence of such an
cryptoexotic baryon.

\subsection*{Acknowledgements}
We would like to thank   J. Haidenbauer, T. Nakano, A. Nogga for useful
discussions.  This work was partially  supported  by Deutsche
Forschungsgemeinschaft  through funds provided to the SFB/TR 16
``Subnuclear Structure of Matter''. This research is part of the \, EU
Integrated \, Infrastructure \, Initiative Hadron Physics Project under
contract  number RII3-CT-2004-506078. A.S. acknowledges support by the
JLab grant SURA-06-C0452  and the 
COSY FFE grant No. 41760632 (COSY-085).

\end{document}